\documentclass[sigconf,natbib]{acmart}

\usepackage{tabularx} 
\usepackage{kotex}
\usepackage{booktabs,multirow}
\usepackage{algorithm}
\usepackage{algpseudocode}
\usepackage{enumitem}
\usepackage{mathtools}

\AtBeginDocument{%
  \providecommand\BibTeX{{%
    \normalfont B\kern-0.5em{\scshape i\kern-0.25em b}\kern-0.8em\TeX}}}

\newcommand{\ie}{{\it i.e.}}

\copyrightyear{2023}
\acmYear{2023}
\setcopyright{acmlicensed}\acmConference[SIGIR '23]{Proceedings of the 46th International ACM SIGIR Conference on Research and Development in Information Retrieval}{July 23--27, 2023}{Taipei, Taiwan}
\acmBooktitle{Proceedings of the 46th International ACM SIGIR Conference on Research and Development in Information Retrieval (SIGIR '23), July 23--27, 2023, Taipei, Taiwan}
\begin{document}

% \title{Unbiased Contrastive Learning via Alignment and Uniformity for Collaborative Filtering}
\title{uCTRL: Unbiased Contrastive Representation Learning via Alignment and Uniformity for Collaborative Filtering}

\author{Jae-woong Lee}
\affiliation{
  \institution{Sungkyunkwan University}
  \country{Republic of Korea}
  }
\email{jwlee.icc@skku.edu}

\author{Seongmin Park}
\affiliation{
  \institution{Sungkyunkwan University}
  \country{Republic of Korea}
  }
\email{psm1206@skku.edu}

\author{Mincheol Yoon}
\affiliation{
  \institution{Sungkyunkwan University}
  \country{Republic of Korea}
  }
\email{yoon56@skku.edu}

\author{Jongwuk Lee}\authornote{Corresponding author}
\affiliation{
  \institution{Sungkyunkwan University}
  \country{Republic of Korea}}
\email{jongwuklee@skku.edu}

\begin{abstract}
Because implicit user feedback for the collaborative filtering (CF) models is biased toward popular items, CF models tend to yield recommendation lists with popularity bias. Previous studies have utilized \emph{inverse propensity weighting (IPW)} or \emph{causal inference} to mitigate this problem. However, they solely employ pointwise or pairwise loss functions and neglect to adopt a contrastive loss function for learning meaningful user and item representations. In this paper, we propose \emph{Unbiased ConTrastive Representation Learning (uCTRL)}, optimizing alignment and uniformity functions derived from the InfoNCE loss function for CF models. Specifically, we formulate an unbiased alignment function used in uCTRL. We also devise a novel IPW estimation method that removes the bias of both users and items. Despite its simplicity, uCTRL equipped with existing CF models consistently outperforms state-of-the-art unbiased recommender models, up to 12.22\% for Recall@20 and 16.33\% for NDCG@20 gains, on four benchmark datasets.
\end{abstract}


 
% , alignment and uniformity

\maketitle

\section{Introduction}\label{sec:Introduction}
Collaborative filtering (CF) is pivotal for building personalized recommender systems. The core idea of CF is based on the \emph{homophily of users}, meaning that users with similar behavior are likely to share similar user preferences. Unlike content-based models using additional features on items, CF models only utilize past user-item interactions to predict hidden user preferences on items. It is thus essential for capturing meaningful collaborative signals.

Because user-item interactions are typically biased toward popular users/items, it is non-trivial to learn unbiased user/item representations. The user-item interaction matrix naturally follows a Pareto data distribution. While a few head items frequently interact with users, many tail items rarely interact with users. CF models are thus easily biased toward popular users/items. As a result, biased user/item representations deviate from \emph{true} user preferences, leading to poor generalization performance. Even worse, they can bring \emph{bias amplification} by over-recommending popular items.

Existing studies~\cite{SaitoYNSN20, Saito20, ZhuHZC20, QinCMNQW20, LeePL21, LeePLL22} utilized new loss functions using \emph{inverse propensity weighting (IPW)} under the \emph{missing-not-at-random (MNAR)} assumption~\cite{MarlinZRS07, Steck10, ZhengGLHLJ21}. Besides, casual inference~\cite{ZhangFHWSLZ21, WeiFCWYH21} was used to alleviate the relationship between true preference and click data. However, they still have several limitations. (i) Existing studies on removing popularity bias in recommender systems have yet to actively incorporate contrastive loss function actively, even though it has been widely used for better representation learning in different domains. (ii) The existing debiasing strategies do not consider users and items when estimating propensity weights, even though recommender models learn latent embedding vectors from user/items interactions.

This paper proposes \emph{Unbiased ConTrastive Representation Learning (uCTRL)}. Motivated by DirectAU~\cite{DirectAU_WangYM000M22}, contrastive representation learning is decomposed into two loss functions: alignment and uniformity. The alignment function represents the distance between user and item vectors for positive user-item interactions. The uniformity function represents the sum of average distances for each user and item distribution. Although DirectAU~\cite{DirectAU_WangYM000M22} shows better performance than existing point-wise and pair-wise loss functions, it is observed that the alignment function is still biased to user/item popularity. To address this issue, we formulate an unbiased alignment function and devise a new method for estimating IPW, debiasing user and item popularity in the alignment function. Extensive experimental results demonstrate that uCTRL equipped with two MF models (\emph{i.e.}, MF~\cite{Koren08} and LightGCN~\cite{0001DWLZ020}) outperforms state-of-the-art unbiased models on four benchmark datasets (\emph{i.e.,} MovieLens 1M, Gowalla, Yelp, and Yahoo! R3).

In summary, the main contributions of this paper are as follows. (1) We introduce a novel loss function for unbiased contrastive representation learning (uCTRL) via alignment and uniformity functions. (2) We also develop a new propensity weighting estimation method to remove the biases of users/items and incorporate it into the unbiased alignment function. (3) We lastly evaluate extensive experiments, and uCTRL outperforms state-of-the-art unbiased CF methods in various unbiased settings.

% % \vspace{-2mm}
% \begin{itemize}
%     \item We first introduce a novel loss function for unbiased contrastive representation learning (uCTRL) via alignment and uniformity functions.
% \end{itemize}
% % \vspace{-2mm}
% \begin{itemize}
%     \item We also develop a new propensity weighting estimation method to remove the biases of users/items and incorporate it into the unbiased alignment function.
% % \vspace{-}
%     \item We lastly evaluate extensive experiments, and uCTRL outperforms state-of-the-art unbiased CF methods in various unbiased settings.
% \end{itemize}

\vspace{-2mm}

\section{Background}\label{sec:background}

\textbf{Notations}. Let $\mathcal{U}$ and $\mathcal{I}$ denote sets of $m$ users and $n$ items, respectively. A user-item interaction matrix $\mathbf{Y} \in \{0, 1\}^{m \times n}$ is given, where $y_{ui}$ is a Bernoulli random variable. If the user $u$ clicks on the item $i$, $y_{ui} = 1$. Otherwise, $y_{ui} = 0$.

CF models aim to predict the true preference for $y_{ui} \in \mathbf{Y}$. To achieve this, existing studies~\cite{SchnabelSSCJ16, ChenDWFWH20, QinCMNQW20, Saito20, SaitoYNSN20, LeePL21, LeePLL22} decompose a user-item click into two components: observation and relevance. The user clicks on an item if (i) the user is aware of the item and (ii) she is also interested in the item.
\begin{equation}\label{eq:click}
P(y_{ui} = 1) = P(o_{ui} = 1) \cdot P(r_{ui} = 1) = \omega_{ui} \cdot \rho_{ui}.
\end{equation}

\noindent
Here, $o_{ui} \in \mathbf{O}$ and $r_{ui} \in \mathbf{R}$ correspond to $y_{ui}$. They represent the Bernoulli random variable for observation and relevance, respectively. For simplicity, let $\omega_{ui}$ and $\rho_{ui}$ denote $P(o_{ui} = 1)$ and $P(r_{ui} = 1)$, respectively.

\vspace{0.5mm}
\noindent
\textbf{Unbiased recommender learning}. The goal of an unbiased ranking function is to debias click data. The loss function for biased feedback is as follows.
\begin{equation}
\label{eq:click_loss}
  \mathcal{L}_\text{biased}(\hat{\textbf{R}}) = \frac{1}{|\mathcal{D}|}\sum_{(u, i) \in \mathcal{D}}
  \left( y_{ui}\delta^{+}_{ui} + \left( 1 - y_{ui} \right) \delta^{-}_{ui} \right),
\end{equation}
where $\mathcal{D}$ denotes the set of all user-item pairs. Additionally, $\delta^{+}_{ui}$ and $\delta^{-}_{ui}$ are the loss function values for the relevant and irrelevant pairs, respectively. Note that cross-entropy loss or mean-square loss functions are commonly used in training CF models.

We define an ideal loss function by replacing click variable $y_{ui}$ with relevance variable $\rho_{ui}$.
\begin{equation}
\label{eq:ideal_loss}
  \mathcal{L}_\text{ideal}(\hat{\textbf{R}}) = \frac{1}{|\mathcal{D}|}\sum_{(u, i) \in \mathcal{D}}
  \left( \rho_{ui}\delta^{+}_{ui} + \left( 1 - \rho_{ui} \right) \delta^{-}_{ui} \right).
\end{equation}

It is vital to bridge the gap between biased and ideal loss functions. Recently, \citet{SaitoYNSN20} adopted the \emph{inverse propensity weighting (IPW)} to remove the bias for click data.
\begin{equation}
\label{eq:saito}
\mathcal{L}_{unbiased}(\hat{\textbf{R}}) = \frac{1}{|\mathcal{D}|}\sum_{(u, i) \in \mathcal{D}}{ \left(\frac{y_{ui}}{\omega_{ui}} \delta^{+}_{ui} + \left ( 1-\frac{y_{ui}}{\omega_{ui}} \right )\delta^{-}_{ui} \right)},
\end{equation}
where $\omega_{ui}$ is a propensity score, meaning the probability that user $u$ observes item $i$. As proved in~\cite{SaitoYNSN20}, the expectation of the unbiased loss function in Eq.~\eqref{eq:saito} is equivalent to the ideal loss function in Eq.~\eqref{eq:ideal_loss}. Therefore, the key issue is how to calculate a propensity score for a user-item interaction pair $(u, i)$.

% ##################################################################################################### 
\vspace{0.5mm}
\noindent
\textbf{Alignment and uniformity}. Recently, \citet{AU_0001I20} have proved that the InfoNCE loss function is optimized by the alignment and uniformity functions. DirectAU~\cite{DirectAU_WangYM000M22} has recently proposed to optimize the two functions for user and item embedding vectors in the recommender system domain. Specifically, the alignment function is defined as the expected distance between the normalized user and item vectors.

\begin{equation}\label{eq:alignmnet}
\mathcal{L}_{\text{align}}(\hat{\textbf{R}})= \mathop{\mathbb{E}}_{(u,i)\sim p_{\text{pos}}} \lVert \tilde{f(u)} - \tilde{f(i)} \rVert^{2},
\end{equation}
where $p_{\text{pos}}$ is the distribution of positive interaction pairs, $f(\cdot) \in \mathbb{R}^d$ means $d$-dimensional user and item vectors, and $\tilde{f(\cdot)}$ means the L2-normalized vector representation. Next, the uniformity function is defined as the sum of the average pairwise Gaussian distribution for users and items.
\begin{align}\label{eq:uniformity}
\mathcal{L}_{\text{uniform}}(\hat{\textbf{R}}) = & \log \mathop{\mathbb{E}}_{u,u'\sim p_{\text{user}}} e^{ -2\lVert \tilde{f(u)} - \tilde{f(u')} \rVert^{2} } /2 \; + \nonumber \\
 & \log \mathop{\mathbb{E}}_{i,i'\sim p_{\text{item}}} e^{ -2\lVert \tilde{f(i)} - \tilde{f(i')} \rVert^{2} } /2,
\end{align}
where $p_{\text{user}}$ and $p_{\text{item}}$ are user and item distributions, respectively. Finally, DirectAU~\cite{DirectAU_WangYM000M22} jointly optimizes the two objective functions.

\begin{equation}\label{eq:DirectAU}
\mathcal{L}_\text{DAU}(\hat{\textbf{R}}) = \mathcal{L}_\text{align}(\hat{\textbf{R}}) + \gamma\mathcal{L}_\text{uniform}(\hat{\textbf{R}}),
\end{equation}
where $\gamma$ is a hyperparameter to control the uniformity loss.

However, it is observed that the alignment is biased for click data. If we consider the distance between user and item vectors as local loss, it is equivalent to training local loss with biased click data as labels, similar to Eq.~\eqref{eq:click_loss}.
\begin{align}\label{eq:alignmnet_bias}
\mathcal{L}_{\text{align}}(\hat{\textbf{R}}) &= \mathop{\mathbb{E}}_{(u,i)\sim p_{\text{pos}}} \lVert \tilde{f(u)} - \tilde{f(i)} \rVert^{2} \nonumber \\
&= \frac{1}{|\mathcal{D}_{click}|}\sum_{(u, i) \in \mathcal{D}_{click}} y_{ui} \lVert \tilde{f(u)} - \tilde{f(i)} \rVert^{2},
\end{align}
where $\mathcal{D}_{click}$ is a set of clicked user-item pairs. It is necessary to optimize the alignment function by eliminating click bias to bring truly relevant user-item pairs closer together in an embedding space.
\vspace{-4mm}
\begin{figure}[t]
\centering
\includegraphics[height=3.8cm]{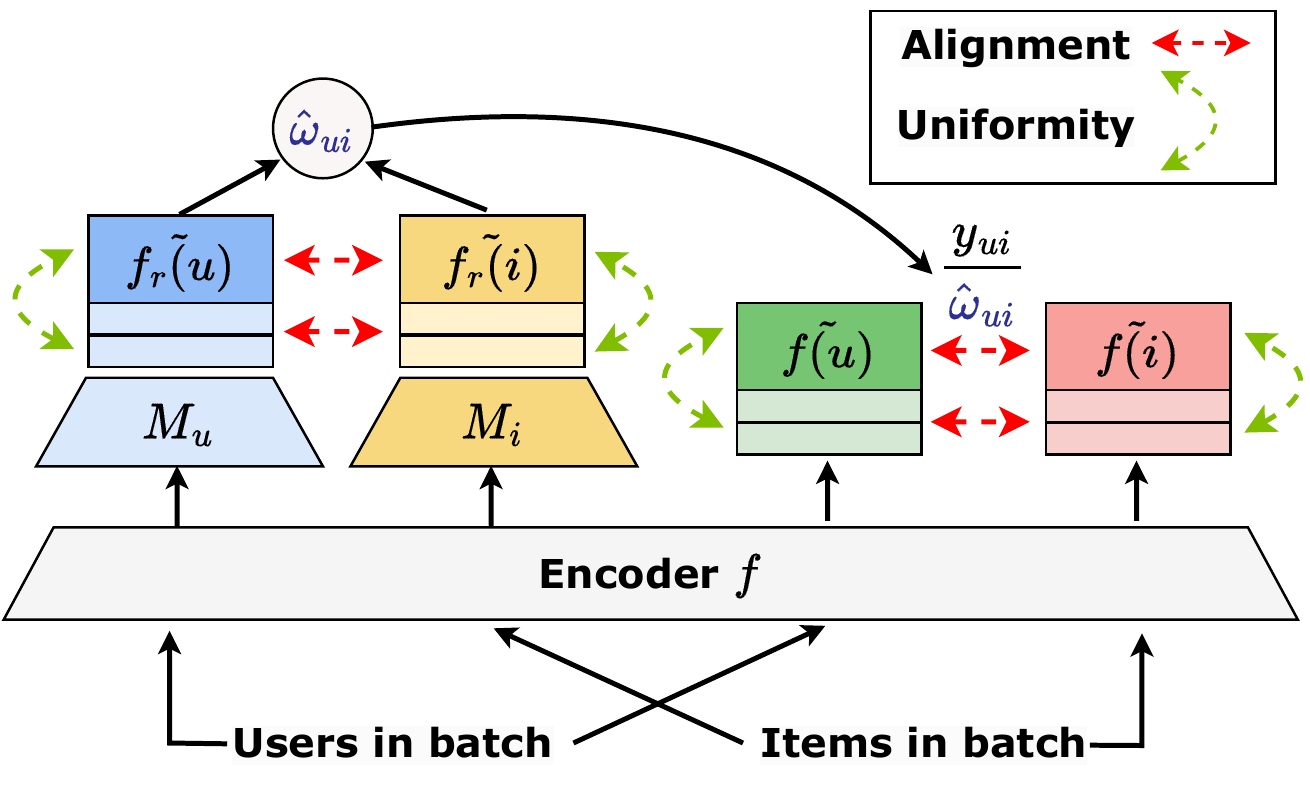}
\vspace{-4mm}

\caption{Overview of the proposed uCTRL. We optimize the alignment loss with propensity weights to remove the bias. Propensity weights are predicted based on the representations of users and items.\vspace{-6.5mm}
}\label{fig:propensity_cal}
\end{figure}

\vspace{-2mm}
\section{Proposed Method}\label{sec:model}

In this section, we present \emph{Unbiased ConTrastive Representation Learning (uCTRL)}, which removes the bias of click data. First, we derive an unbiased alignment loss to debias click data. Then, we propose a novel method for estimating the propensity weight $\omega_{ui}$ used in the unbiased alignment function. Finally, we jointly learn the unbiased DirectAU and the propensity weight estimation model.

\vspace{0.5mm}
\noindent
\textbf{Unbiased DirectAU}. While DirectAU~\cite{DirectAU_WangYM000M22} is effective in training user and item representations via contrastive learning; however, it is observed that the alignment loss is biased to click data as in Eq.~\eqref{eq:alignmnet_bias}. This is different from an ideal loss function via true user preferences. Given a set of clicked user-item pairs $\mathcal{D}_{click}$, the ideal alignment loss for relevance is formulated as follows.
\begin{equation}\label{eq:alignmnet_ideal}
\mathcal{L}_{\text{ideal\_align}}(\hat{\textbf{R}}) =\frac{1}{|\mathcal{D}_{click}|}\sum_{(u, i) \in \mathcal{D}_{click}} \rho_{ui} \lVert \tilde{f(u)} - \tilde{f(i)} \rVert^{2}.
\end{equation}

To reduce the gap between ideal and biased alignment loss functions, we introduce IPW as used in existing studies~\cite{SaitoYNSN20, Saito20, ZhuHZC20, LeePL21, LeePLL22}.
\begin{equation}\label{eq:unbiased_alignmnet}
\mathcal{L}_{\text{unbiased\_align}}(\hat{\textbf{R}}) =\frac{1}{|\mathcal{D}_{click}|}\sum_{(u, i) \in \mathcal{D}_{click}} \frac{y_{ui}}{\omega_{ui}} \lVert \tilde{f(u)} - \tilde{f(i)} \rVert^{2}.
\end{equation}

Despite its simplicity, it is expected to converge to the ideal alignment loss for true user preferences.
\begin{align}
  \mathbb{E}\left [ \mathcal{L}_\text{unbiased\_align}(\hat{\textbf{R}}) \right ] &=\frac{1}{|\mathcal{D}_{click}|}\sum_{(u, i) \in \mathcal{D}_{click}} \frac{\mathbb{E}\left [y_{ui}\right ]}{\omega_{ui}} \lVert \tilde{f(u)} - \tilde{f(i)} \rVert^{2}  \nonumber \\
  &=\frac{1}{|\mathcal{D}_{click}|}\sum_{(u, i) \in \mathcal{D}_{click}} \frac{\omega_{ui} \rho_{ui}}{\omega_{ui}} \lVert \tilde{f(u)} - \tilde{f(i)} \rVert^{2}  \nonumber \\
  & = \frac{1}{|\mathcal{D}_{click}|}\sum_{(u, i) \in \mathcal{D}_{click}} \rho_{ui} \lVert \tilde{f(u)} - \tilde{f(i)} \rVert^{2}.
\end{align}
\vspace{-3mm}

Unlike the alignment loss, the original uniformity loss utilizes sampled user and item pairs regardless of the bias of click data. It does not require the debiasing method for computing user and item uniformity losses. Finally, the unbiased DirectAU approach optimizes unbiased representations for positive user-item pairs by aligning them with true preference while also promoting uniformity for both users and items.
\begin{equation}\label{eq:unbiased_DirectAU}
\mathcal{L}_\text{unbiased\_DAU}(\hat{\textbf{R}}) = \mathcal{L}_{\text{unbiased\_align}}(\hat{\textbf{R}}) + \gamma\mathcal{L}_\text{uniform}(\hat{\textbf{R}}).
\end{equation}

\noindent
\textbf{Calculating propensity weights}. The key issue for computing the unbiased alignment loss is to estimate the propensity weights for click data. While the propensity weights for user-item interactions heavily affect the popularity of both users and items, existing methods~\cite{SaitoYNSN20, Saito20, ZhuHZC20, LeePL21, LeePLL22} mostly focus on estimating item popularity. To address this limitation, we adopt a new method that incorporates both user and item vectors for predicting the propensity weights.

Specifically, we borrow the key idea of TransR~\cite{TransR_LinLSLZ15}, which projects different entities on the entity space into relation space through a projection matrix. Then, it calculates the entity relationship in the relationship space. Although user and item vectors are different information, they are related. Therefore, we first transform the original user and item vectors into the relation space to compute the propensity weights.
\begin{equation}\label{eq:propensity_vector}
f_r(u) = \tilde{f(u)}\mathbf{M}_{u}^{\top}, f_r(i) = \tilde{f(i)}\mathbf{M}_{i}^{\top},
\end{equation}
where $\mathbf{M}_u \in \mathbb{R}^{d \times d}$ and $\mathbf{M}_i \in \mathbb{R}^{d \times d}$ are the projection matrices that reflect the different characteristics of users and items. $f_r(u)$ and $f_r(i)$ are the projected vectors of user $u$ and item $i$.

To learn the projection matrices, we utilize DirectAU with click data. By replacing $\tilde{f(u)}$ and $\tilde{f(i)}$ with $\tilde{f_r(u)}$ and $\tilde{f_r(i)}$ in Eq.~\eqref{eq:alignmnet} and Eq.~\eqref{eq:uniformity}, we optimize alignment and uniformity as follows:
\begin{equation}\label{eq:final_uCTRL}
\mathcal{L}_\text{relation\_DAU}(\hat{\textbf{R}}) = \mathcal{L}_{\text{relation\_align}}(\hat{\textbf{R}}) + \lambda\mathcal{L}_\text{relation\_uniform}(\hat{\textbf{R}}),
\end{equation}
where $\lambda$ is a hyperparameter to adjust the uniformity loss. Note that we have frozen $\tilde{f(u)}$ and $\tilde{f(i)}$ within $\tilde{f_r(u)}$ and $\tilde{f_r(i)}$ to exclusively train the projection matrices.

Because users and items are represented in the relationship space, we calculate the propensity weights for $\omega_{ui}$ by taking the dot product of the two vectors.
\begin{equation}\label{eq:propensity_inference}
\hat{\omega}_{ui} = \sigma \left(\tilde{f_r(u)} \cdot \tilde{f_r(i)}^{\top} \right),
\end{equation}

\noindent
where $\hat{\omega}_{ui}$ is the estimated propensity weights, $\sigma \left( \cdot \right)$ is the sigmoid function, and $\tilde{f_r(\cdot)}$ is the L2-normalized projected vector. Finally, the propensity score is estimated to remove both the bias of users and items.

To handle the high variance problem of IPW, we also apply the clipping method, commonly used in the variance reduction technique. We restrict the minimum value of $\hat{\omega}_{ui}$.
\begin{equation}\label{eq:propensity_clipping}
\hat{\omega}_{ui} = \text{max}\{\hat{\omega}_{ui}, \mu \},
\end{equation}
\noindent
where $\mu$ is the hyperparameter. Empirically, we set $\mu$ to 0.1.

\vspace{0.5mm}
\noindent
\textbf{uCTRL loss function}. The final loss function of uCTRL consists of two parts: the unbiased DirectAU loss to learn true preferences and the propensity weight estimation loss to learn projected vectors of users and items in the relationship space. It jointly learns two loss functions for the original and relation spaces.
\begin{equation}\label{eq:final_uCTRL}
\mathcal{L}_\text{uCTRL}(\hat{\textbf{R}}) = \mathcal{L}_\text{unbiased\_DAU}(\hat{\textbf{R}}) + \mathcal{L}_\text{relation\_DAU}(\hat{\textbf{R}}).
\end{equation}

\section{EXPERIMENTS}

\subsection{Experimental Setup}
\label{sec:setup}

\begin{table*}[t]
\centering
\caption{
Comparative results on four datasets averaged over 5 runs (* and ** indicating $p < 0.05$ and $p < 0.01$ for a one-tailed t-test with DirectAU). The \textit{Gain} indicates the improvement ratio over DirectAU. The best models are highlighted in bold font.}\label{tab:main_table}

\vspace{-4.5mm}

\begin{center}
\renewcommand{\arraystretch}{0.68} % 테이블 행 간격
% \begin{tabularx}{\textwidth}{p{.12\textwidth} c{.12\textwidth} c{.12\textwidth} | c{.12\textwidth} c{.12\textwidth} c{.12\textwidth} c{.12\textwidth} c{.12\textwidth} c{.12\textwidth} c{.12\textwidth} | c{.12\textwidth} | c{.12\textwidth}}
{\small
\begin{tabular}{c c c | c c c c c c c | c | c}
% \begin{tabular}{c{.12\textwidth} c{.12\textwidth} c{.12\textwidth} | c{.12\textwidth} c{.12\textwidth} c{.12\textwidth} c{.12\textwidth} c{.12\textwidth} c{.12\textwidth} c{.12\textwidth} | c{.12\textwidth} | c{.12\textwidth}}
\toprule
Dataset & Backbone & Metric & Rel & SIPW & PDA & MACR & CCL &  BCLoss & DirectAU & uCTRL & \textit{Gain (\%)} \\
\hline
\multirow{4}{*}{ML-1M}
& \multirow{2}{*}{MF}
& Recall@20 & 0.1285 & 0.1256 & 0.1424 & 0.1558 & 0.0919 & 0.1498 & 0.1539 & \textbf{0.1727**} & \textit{+12.22} \\
& & NDCG@20 & 0.1046 & 0.1014 & 0.1152 & 0.1253 & 0.0705 & 0.1264 & 0.1139 & \textbf{0.1325**} & \textit{+16.33} \\
\cline{2-12}
& \multirow{2}{*}{LightGCN}
& Recall@20 & 0.0754 & 0.0814 & 0.0652 & 0.0413 & 0.0862 & 0.1233 & 0.1633 & \textbf{0.1707**} & \textit{+4.53} \\
& & NDCG@20 & 0.0550 & 0.0514 & 0.0627 & 0.0403 & 0.0633 & 0.1025 & 0.1230 & \textbf{0.1319**} & \textit{+7.24} \\
\hline
\multirow{4}{*}{Gowalla}
& \multirow{2}{*}{MF}
& Recall@20 & 0.0771 & 0.0417 & 0.0655 & 0.0704 & 0.0918 & 0.0567 & 0.1213 & \textbf{0.1265**} & \textit{+4.29} \\
& & NDCG@20 & 0.0431 & 0.0221 & 0.0337 & 0.0382 & 0.0531 & 0.0304 & 0.0718 & \textbf{0.0742**} & \textit{+3.34} \\
\cline{2-12}
& \multirow{2}{*}{LightGCN}
& Recall@20 & 0.0458 & 0.0475 & 0.0423 & 0.0218 & 0.0698 & 0.0335 & 0.0960 & \textbf{0.1007**} & \textit{+4.88} \\
& & NDCG@20 & 0.0227 & 0.0244 & 0.0198 & 0.0112 & 0.0370 & 0.0165 & 0.0547 & \textbf{0.0567**} & \textit{+3.62} \\
\hline
\multirow{4}{*}{Yelp}
& \multirow{2}{*}{MF}
& Recall@20 & 0.0253 & 0.0151 & 0.0228 & 0.0256 & 0.0047 & 0.0242 & 0.0480 & \textbf{0.0508**} & \textit{+5.83} \\
& & NDCG@20 & 0.0153 & 0.0091 & 0.0132 & 0.0155 & 0.0028 & 0.0146 & 0.0304 & \textbf{0.0321**} & \textit{+5.59} \\
\cline{2-12}
& \multirow{2}{*}{LightGCN}
& Recall@20 & 0.0126 & 0.0175 & 0.0199 & 0.0040 & 0.0150 & 0.0146 & 0.0423 & \textbf{0.0441*\;} & \textit{+4.26} \\
& & NDCG@20 & 0.0064 & 0.0103 & 0.0116 & 0.0022 & 0.0089 & 0.0083 & 0.0262 & \textbf{0.0272*\;} & \textit{+3.82} \\
\hline
\hline

\multirow{4}{*}{Yahoo! R3}
& \multirow{2}{*}{MF}
& Recall@20 &  0.0288 & 0.0278 & 0.0286 & 0.0294 & 0.0269 & 0.0290 & 0.0281 & \textbf{0.0297*\;} & \textit{+5.58} \\ 
& & NDCG@20 &  0.0188 & 0.0177 & 0.0194 & \textbf{0.0199} & 0.0176 & 0.0198 & 0.0186 & \textbf{0.0199**} & \textit{+7.32} \\
\cline{2-12}
& \multirow{2}{*}{LightGCN}
& Recall@20 &  0.0262 & 0.0255 & 0.0280 & 0.0284 & 0.0275 & 0.0296 & 0.0297 & \textbf{0.0300\;\;} & \textit{+1.00} \\
& & NDCG@20 &  0.0173 & 0.0165 & 0.0185 & 0.0190 & 0.0179 & 0.0198 & 0.0202 & \textbf{0.0203\;\;} & \textit{+0.83} \\

\bottomrule
\end{tabular}
}
\end{center}
% \vspace{-5mm}
\end{table*}

\begin{table}[t]
\centering

\caption{Comparison of the propensity weights for the "+Rel", "+SIPW", and proposed method using uCTRL on ML-1M and Yahoo! R3 datasets. We select ML-1M as the representative one for synthetic datasets. Each metric is @20.}\label{tab:propensity_effect} \small
% \vspace{-4.5mm}
\vspace{-4mm}

\begin{center}
\renewcommand{\arraystretch}{0.65} % 테이블 행 간격
{\small
\begin{tabular}{c c | c c c c}
\toprule

\multicolumn{2}{c}{} & \multicolumn{2}{c}{ML-1M} & \multicolumn{2}{c}{Yahoo! R3} \\
\cmidrule(lr){3-4}
\cmidrule(lr){5-6}
Backbone & Method & Recall & NDCG & Recall & NDCG \\
\hline
\multirow{3}{*}{MF}
 & uCTRL & \textbf{0.1727} & \textbf{0.1325} & \textbf{0.0297} & \textbf{0.0199} \\
 & + Rel & 0.0963 & 0.0749 & 0.0219 & 0.0131 \\
 & + SIPW & 0.1431 & 0.1057 & 0.0282 & 0.0182 \\
 \hline
\multirow{3}{*}{LightGCN}
 & uCTRL & \textbf{0.1707} & \textbf{0.1319} & \textbf{0.0300} & \textbf{0.0203} \\
 & + Rel & 0.0715 & 0.0599 & 0.0270 & 0.0174 \\
 & + SIPW & 0.1167 & 0.0827 & 0.0293 & 0.0202 \\
\bottomrule
\end{tabular}
}
\end{center}
\vspace{-3.8mm}
\vspace{-3mm}
\end{table}

% \vspace{2mm}
% \noindent
\textbf{Datasets}.
We evaluate our proposed uCTRL with the following four datasets: ML-1M\footnote{\vspace{-0.6mm}\url{http://grouplens.org/datasets/movielens/}} (6K users, 3K items, and 574K interactions), Gowalla\footnote{\vspace{-0.6mm}\url{http://snap.stanford.edu/data/loc-gowalla.html}} (30K users, 41K items, and 925K interactions), Yelp\footnote{\vspace{-0.6mm}\url{https://www.yelp.com/dataset}} (32K users, 38K items, and 1,405K interactions), and Yahoo! R3\footnote{\vspace{-0.6mm}\url{http://webscope.sandbox.yahoo.com/}} (15K users, 1K items, and 366K interactions).

\vspace{1mm}
\noindent
\textbf{Preprocessing}. 
For data preprocessing, we categorized the four datasets into two types based on the manner in which the test data was collected. (1) Synthetic debiased test data (ML-1M, Gowalla, and Yelp): This dataset comprises biased click data. To evaluate the effect of bias removal, we followed the approach used in previous studies~\cite{VasileB19, WangLCB20, ZhengGLHLJ21, WeiFCWYH21}. It involves uniformly sampling the test set across all items. We split 10\% as an unbiased test set, 10\% as a biased validation set, and the remaining 80\% as a biased training set. (2) True debiased test data (Yahoo! R3): This training data was collected with bias, and the test data was collected without bias through a survey. We followed the data preprocessing of Recbole-Debias\footnote{\vspace{-0.5mm}\url{https://github.com/JingsenZhang/Recbole-Debias/}}. The code and detailed hyperparameter settings can be accessed at \url{https://github.com/Jaewoong-Lee/sigir\_2023\_uCTRL}.

\vspace{1mm}
\noindent
\textbf{Competitive methods}. We compare our proposed method with recent approaches in unbiased and contrastive learning, which aim to mitigate popularity bias by selecting from existing methods~\cite{SaitoYNSN20, Saito20, ZhuHZC20, QinCMNQW20, LeePL21, LeePLL22, MaoZWDDXH21, bc_loss, DirectAU_WangYM000M22, liu2021contrastive, abs-2302-08191, wang2023knowledge}. For unbiased methods, we compare it with Rel~\cite{SaitoYNSN20} and SIPW~\cite{LeePLL22}, both IPW-based methods. SIPW utilizes the self-inverse propensity weighting of \cite{LeePLL22}. PDA~\cite{ZhangFHWSLZ21} and MACR~\cite{WeiFCWYH21} are methods that use causality to remove bias. Especially, MACR considers both item popularity and user conformity bias. For contrastive learning methods, we bring CCL~\cite{MaoZWDDXH21}, which uses hard negative filtering with a margin, and BCLoss~\cite{bc_loss}, a method for mitigating popularity bias based on InfoNCE loss. Besides, we introduce DirectAU~\cite{DirectAU_WangYM000M22}, which optimizes alignment and uniformity. We adopt MF~\cite{Koren08} and LightGCN~\cite{0001DWLZ020} as backbones for all the models.

% \vspace{1mm}
\noindent
\textbf{Evaluation metrics}. We use two evaluation metrics: NDCG and recall. Given the importance of accurately recommending top-ranked items in a top-N recommendation scenario, top-N recommendation metrics are computed based on the highest 20 ranked items $\left ( @20 \right)$ for all the datasets.

\vspace{-2mm}
\subsection{Experimental Results}
\label{sec:result}

\begin{table}[t]
\centering

\caption{Comparison of alignment loss value between DirectAU and uCTRL on ML-1M. All the users (\ie,~$u$) and items (\ie,~$i$) are split into popular (\ie,~Pop.) and unpopular (\ie,~Unpop.) groups with a ratio of 2:8 based on popularity. The \textit{Drop} indicates the decrease ratio of alignment loss over DirectAU.}\label{tab:align_by_pop} \small
\vspace{-4.5mm}
% \vspace{-4.5mm}
\begin{center}
{\small
\renewcommand{\arraystretch}{0.65} % 테이블 행 간격
\begin{tabular}{c c | c c | c c }
\toprule
Backbone & Model & Pop. $u$ & Unpop. $u$ & Pop. $i$ & Unpop. $i$ \\
\hline
\multirow{3}{*}{MF}
& DirectAU & 1.6370 & 1.6071 & 1.6836 & 1.5021 \\
& uCTRL & 1.5696 & 1.5396 & 1.6173 & 1.4327 \\
\cline{2-6}
& \textit{Drop (\%)} & \textit{-4.12} & \textit{-4.20} & \textit{-3.94} & \textit{-4.62} \\
\hline
\multirow{3}{*}{LightGCN}
& DirectAU & 1.7223 & 1.6934 & 1.7666 & 1.5935 \\
& uCTRL & 1.6745 & 1.6429 & 1.7224 & 1.5347 \\
\cline{2-6}
& \textit{Drop (\%)} & \textit{-2.78} & \textit{-2.98} & \textit{-2.51} & \textit{-3.69} \\
\bottomrule
\end{tabular}
}
\end{center}
\vspace{-5.8mm}
\end{table}

\noindent
\textbf{Comparison with baselines.} Table~\ref{tab:main_table} presents a performance comparison between our proposed method and the other baselines. For all the datasets, Our proposed uCTRL consistently outperforms existing models on two backbones, MF and LightGCN. Notably, our proposed uCTRL beats DirectAU up to 16.33\%, 4.88\%, 5.83\%, and 7.32\% on the ML-1M, Gowalla, Yelp, and Yahoo! R3 datasets, respectively. These results show that our proposed method effectively learns unbiased representations of users and items, leading to improved performance.

\vspace{0.5mm}
\noindent
\textbf{Effect of our proposed propensity weight.} To assess the debiasing effectiveness of our proposed propensity weight, we compare it with two other propensity weights, Rel~\cite{SaitoYNSN20} and SIPW~\cite{LeePLL22}. We replace Rel and SIPW with our proposed propensity weight in uCTRL. Table~\ref{tab:propensity_effect} indicates that our proposed propensity weights perform better than the other two methods in reducing bias.

\vspace{0.5mm}
\noindent
\textbf{Comparison of alignment loss between DirectAU and uCTRL.} Table~\ref{tab:align_by_pop} illustrates the difference in the alignment loss between DirectAU and uCTRL. We computed the alignment loss using user and item vectors in Eq.~\eqref{eq:alignmnet} and divided users and items into two groups based on their popularity. Based on the results, we observed two interesting findings. First, the decrease in the unpopular group was greater than that in the popular group when comparing Pop. $u$ vs. Unpop. $u$ and Pop. $i$ vs. Unpop. $i$. This indicates that the proposed method has a more effective debiasing effect on the unpopular group. Secondly, uCTRL had a lower alignment loss in all the groups compared to DirectAU, implying that uCTRL improves not only the performance of the unpopular groups but also the popular groups.

\vspace{-2mm}
\section{Conclusion}\label{sec:conclusion}
In this paper, we propose \emph{Unbiased ConTrastive Representation Learning (uCTRL)}, which directly optimizes the unbiased alignment and uniformity losses for better representation learning and estimates the propensity scores considering both users and items. To the best of our knowledge, this is the first work that eliminates bias with alignment and uniformity. Our experiments on four real-world datasets confirm that \emph{uCTRL} effectively removes bias and improves ranking performance, outperforming state-of-the-art unbiased recommender models and models using contrastive loss.

% \vspace{-2mm}
% \section*{Acknowledgment}
% This work was supported by Institute of Information \& communications Technology Planning \& Evaluation (IITP) grant funded by the Korea government (MSIT) (No. 2022-0-00680, 2022-0-01045, 2019-0-00421, 2021-0-02068, and IITP-2023-2020-0-01821).
% \vspace{-2mm}

\bibliographystyle{ACM-Reference-Format}
\bibliography{references}


\begin{thebibliography}{25}

%%% the Web address.

\ifx \showCODEN    \undefined \def \showCODEN     #1{\unskip}     \fi
\ifx \showDOI      \undefined \def \showDOI       #1{#1}\fi
\ifx \showISBNx    \undefined \def \showISBNx     #1{\unskip}     \fi
\ifx \showISBNxiii \undefined \def \showISBNxiii  #1{\unskip}     \fi
\ifx \showISSN     \undefined \def \showISSN      #1{\unskip}     \fi
\ifx \showLCCN     \undefined \def \showLCCN      #1{\unskip}     \fi
\ifx \shownote     \undefined \def \shownote      #1{#1}          \fi
\ifx \showarticletitle \undefined \def \showarticletitle #1{#1}   \fi
\ifx \showURL      \undefined \def \showURL       {\relax}        \fi
% invisible to TeX
\providecommand\bibfield[2]{#2}
\providecommand\bibinfo[2]{#2}
\providecommand\natexlab[1]{#1}
\providecommand\showeprint[2][]{arXiv:#2}

\bibitem[Cai et~al\mbox{.}(2023)]%
        {abs-2302-08191}
\bibfield{author}{\bibinfo{person}{Xuheng Cai}, \bibinfo{person}{Chao Huang},
  \bibinfo{person}{Lianghao Xia}, {and} \bibinfo{person}{Xubin Ren}.}
  \bibinfo{year}{2023}\natexlab{}.
\newblock \showarticletitle{LightGCL: Simple Yet Effective Graph Contrastive
  Learning for Recommendation}.
\newblock \bibinfo{journal}{\emph{CoRR}}  \bibinfo{volume}{abs/2302.08191}
  (\bibinfo{year}{2023}).
\newblock


\bibitem[Chen et~al\mbox{.}(2020)]%
        {ChenDWFWH20}
\bibfield{author}{\bibinfo{person}{Jiawei Chen}, \bibinfo{person}{Hande Dong},
  \bibinfo{person}{Xiang Wang}, \bibinfo{person}{Fuli Feng},
  \bibinfo{person}{Meng Wang}, {and} \bibinfo{person}{Xiangnan He}.}
  \bibinfo{year}{2020}\natexlab{}.
\newblock \showarticletitle{Bias and Debias in Recommender System: {A} Survey
  and Future Directions}.
\newblock \bibinfo{journal}{\emph{CoRR}}  \bibinfo{volume}{abs/2010.03240}
  (\bibinfo{year}{2020}).
\newblock


\bibitem[He et~al\mbox{.}(2020)]%
        {0001DWLZ020}
\bibfield{author}{\bibinfo{person}{Xiangnan He}, \bibinfo{person}{Kuan Deng},
  \bibinfo{person}{Xiang Wang}, \bibinfo{person}{Yan Li},
  \bibinfo{person}{Yong{-}Dong Zhang}, {and} \bibinfo{person}{Meng Wang}.}
  \bibinfo{year}{2020}\natexlab{}.
\newblock \showarticletitle{LightGCN: Simplifying and Powering Graph
  Convolution Network for Recommendation}. In
  \bibinfo{booktitle}{\emph{SIGIR}}. \bibinfo{pages}{639--648}.
\newblock


\bibitem[Koren(2008)]%
        {Koren08}
\bibfield{author}{\bibinfo{person}{Yehuda Koren}.}
  \bibinfo{year}{2008}\natexlab{}.
\newblock \showarticletitle{Factorization meets the neighborhood: a
  multifaceted collaborative filtering model}. In
  \bibinfo{booktitle}{\emph{KDD}}. \bibinfo{pages}{426--434}.
\newblock


\bibitem[Lee et~al\mbox{.}(2021)]%
        {LeePL21}
\bibfield{author}{\bibinfo{person}{Jae{-}woong Lee}, \bibinfo{person}{Seongmin
  Park}, {and} \bibinfo{person}{Jongwuk Lee}.} \bibinfo{year}{2021}\natexlab{}.
\newblock \showarticletitle{Dual Unbiased Recommender Learning for Implicit
  Feedback}. In \bibinfo{booktitle}{\emph{{SIGIR}}}.
  \bibinfo{pages}{1647--1651}.
\newblock


\bibitem[Lee et~al\mbox{.}(2022)]%
        {LeePLL22}
\bibfield{author}{\bibinfo{person}{Jae{-}woong Lee}, \bibinfo{person}{Seongmin
  Park}, \bibinfo{person}{Joonseok Lee}, {and} \bibinfo{person}{Jongwuk Lee}.}
  \bibinfo{year}{2022}\natexlab{}.
\newblock \showarticletitle{Bilateral Self-unbiased Learning from Biased
  Implicit Feedback}. In \bibinfo{booktitle}{\emph{{SIGIR}}}.
  \bibinfo{publisher}{{ACM}}, \bibinfo{pages}{29--39}.
\newblock


\bibitem[Lin et~al\mbox{.}(2015)]%
        {TransR_LinLSLZ15}
\bibfield{author}{\bibinfo{person}{Yankai Lin}, \bibinfo{person}{Zhiyuan Liu},
  \bibinfo{person}{Maosong Sun}, \bibinfo{person}{Yang Liu}, {and}
  \bibinfo{person}{Xuan Zhu}.} \bibinfo{year}{2015}\natexlab{}.
\newblock \showarticletitle{Learning Entity and Relation Embeddings for
  Knowledge Graph Completion}. In \bibinfo{booktitle}{\emph{{AAAI}}}.
  \bibinfo{publisher}{{AAAI} Press}, \bibinfo{pages}{2181--2187}.
\newblock


\bibitem[Liu et~al\mbox{.}(2021)]%
        {liu2021contrastive}
\bibfield{author}{\bibinfo{person}{Zhuang Liu}, \bibinfo{person}{Yunpu Ma},
  \bibinfo{person}{Yuanxin Ouyang}, {and} \bibinfo{person}{Zhang Xiong}.}
  \bibinfo{year}{2021}\natexlab{}.
\newblock \showarticletitle{Contrastive learning for recommender system}.
\newblock \bibinfo{journal}{\emph{arXiv preprint arXiv:2101.01317}}
  (\bibinfo{year}{2021}).
\newblock


\bibitem[Mao et~al\mbox{.}(2021)]%
        {MaoZWDDXH21}
\bibfield{author}{\bibinfo{person}{Kelong Mao}, \bibinfo{person}{Jieming Zhu},
  \bibinfo{person}{Jinpeng Wang}, \bibinfo{person}{Quanyu Dai},
  \bibinfo{person}{Zhenhua Dong}, \bibinfo{person}{Xi Xiao}, {and}
  \bibinfo{person}{Xiuqiang He}.} \bibinfo{year}{2021}\natexlab{}.
\newblock \showarticletitle{SimpleX: {A} Simple and Strong Baseline for
  Collaborative Filtering}. In \bibinfo{booktitle}{\emph{{CIKM}}}.
  \bibinfo{publisher}{{ACM}}, \bibinfo{pages}{1243--1252}.
\newblock


\bibitem[Marlin et~al\mbox{.}(2007)]%
        {MarlinZRS07}
\bibfield{author}{\bibinfo{person}{Benjamin~M. Marlin},
  \bibinfo{person}{Richard~S. Zemel}, \bibinfo{person}{Sam~T. Roweis}, {and}
  \bibinfo{person}{Malcolm Slaney}.} \bibinfo{year}{2007}\natexlab{}.
\newblock \showarticletitle{Collaborative Filtering and the Missing at Random
  Assumption}. In \bibinfo{booktitle}{\emph{UAI}}. \bibinfo{pages}{267--275}.
\newblock


\bibitem[Qin et~al\mbox{.}(2020)]%
        {QinCMNQW20}
\bibfield{author}{\bibinfo{person}{Zhen Qin}, \bibinfo{person}{Suming~J. Chen},
  \bibinfo{person}{Donald Metzler}, \bibinfo{person}{Yongwoo Noh},
  \bibinfo{person}{Jingzheng Qin}, {and} \bibinfo{person}{Xuanhui Wang}.}
  \bibinfo{year}{2020}\natexlab{}.
\newblock \showarticletitle{Attribute-Based Propensity for Unbiased Learning in
  Recommender Systems: Algorithm and Case Studies}. In
  \bibinfo{booktitle}{\emph{KDD}}. \bibinfo{pages}{2359–2367}.
\newblock


\bibitem[Saito(2020)]%
        {Saito20}
\bibfield{author}{\bibinfo{person}{Yuta Saito}.}
  \bibinfo{year}{2020}\natexlab{}.
\newblock \showarticletitle{Unbiased Pairwise Learning from Biased Implicit
  Feedback}. In \bibinfo{booktitle}{\emph{ICTIR}}. \bibinfo{pages}{5--12}.
\newblock


\bibitem[Saito et~al\mbox{.}(2020)]%
        {SaitoYNSN20}
\bibfield{author}{\bibinfo{person}{Yuta Saito}, \bibinfo{person}{Suguru
  Yaginuma}, \bibinfo{person}{Yuta Nishino}, \bibinfo{person}{Hayato Sakata},
  {and} \bibinfo{person}{Kazuhide Nakata}.} \bibinfo{year}{2020}\natexlab{}.
\newblock \showarticletitle{Unbiased Recommender Learning from
  Missing-Not-At-Random Implicit Feedback}. In
  \bibinfo{booktitle}{\emph{WSDM}}. \bibinfo{pages}{501--509}.
\newblock


\bibitem[Schnabel et~al\mbox{.}(2016)]%
        {SchnabelSSCJ16}
\bibfield{author}{\bibinfo{person}{Tobias Schnabel}, \bibinfo{person}{Adith
  Swaminathan}, \bibinfo{person}{Ashudeep Singh}, \bibinfo{person}{Navin
  Chandak}, {and} \bibinfo{person}{Thorsten Joachims}.}
  \bibinfo{year}{2016}\natexlab{}.
\newblock \showarticletitle{Recommendations as Treatments: Debiasing Learning
  and Evaluation}. In \bibinfo{booktitle}{\emph{{ICML}}}
  \emph{(\bibinfo{series}{{JMLR} Workshop and Conference Proceedings},
  Vol.~\bibinfo{volume}{48})}. \bibinfo{pages}{1670--1679}.
\newblock


\bibitem[Steck(2010)]%
        {Steck10}
\bibfield{author}{\bibinfo{person}{Harald Steck}.}
  \bibinfo{year}{2010}\natexlab{}.
\newblock \showarticletitle{Training and testing of recommender systems on data
  missing not at random}. In \bibinfo{booktitle}{\emph{KDD}}.
  \bibinfo{pages}{713--722}.
\newblock


\bibitem[Vasile and Bonner(2019)]%
        {VasileB19}
\bibfield{author}{\bibinfo{person}{Flavian Vasile} {and}
  \bibinfo{person}{Stephen Bonner}.} \bibinfo{year}{2019}\natexlab{}.
\newblock \showarticletitle{Causal Embeddings for Recommendation: An Extended
  Abstract}. In \bibinfo{booktitle}{\emph{{IJCAI}}}.
  \bibinfo{publisher}{ijcai.org}, \bibinfo{pages}{6236--6240}.
\newblock


\bibitem[Wang et~al\mbox{.}(2022)]%
        {DirectAU_WangYM000M22}
\bibfield{author}{\bibinfo{person}{Chenyang Wang}, \bibinfo{person}{Yuanqing
  Yu}, \bibinfo{person}{Weizhi Ma}, \bibinfo{person}{Min Zhang},
  \bibinfo{person}{Chong Chen}, \bibinfo{person}{Yiqun Liu}, {and}
  \bibinfo{person}{Shaoping Ma}.} \bibinfo{year}{2022}\natexlab{}.
\newblock \showarticletitle{Towards Representation Alignment and Uniformity in
  Collaborative Filtering}. In \bibinfo{booktitle}{\emph{{KDD}}}.
  \bibinfo{publisher}{{ACM}}, \bibinfo{pages}{1816--1825}.
\newblock


\bibitem[Wang et~al\mbox{.}(2023)]%
        {wang2023knowledge}
\bibfield{author}{\bibinfo{person}{Hao Wang}, \bibinfo{person}{Yao Xu},
  \bibinfo{person}{Cheng Yang}, \bibinfo{person}{Chuan Shi},
  \bibinfo{person}{Xin Li}, \bibinfo{person}{Ning Guo}, {and}
  \bibinfo{person}{Zhiyuan Liu}.} \bibinfo{year}{2023}\natexlab{}.
\newblock \showarticletitle{Knowledge-Adaptive Contrastive Learning for
  Recommendation}. In \bibinfo{booktitle}{\emph{Proceedings of the Sixteenth
  ACM International Conference on Web Search and Data Mining}}.
  \bibinfo{pages}{535--543}.
\newblock


\bibitem[Wang and Isola(2020)]%
        {AU_0001I20}
\bibfield{author}{\bibinfo{person}{Tongzhou Wang} {and}
  \bibinfo{person}{Phillip Isola}.} \bibinfo{year}{2020}\natexlab{}.
\newblock \showarticletitle{Understanding Contrastive Representation Learning
  through Alignment and Uniformity on the Hypersphere}. In
  \bibinfo{booktitle}{\emph{{ICML}}} \emph{(\bibinfo{series}{Proceedings of
  Machine Learning Research}, Vol.~\bibinfo{volume}{119})}.
  \bibinfo{publisher}{{PMLR}}, \bibinfo{pages}{9929--9939}.
\newblock


\bibitem[Wang et~al\mbox{.}(2020)]%
        {WangLCB20}
\bibfield{author}{\bibinfo{person}{Yixin Wang}, \bibinfo{person}{Dawen Liang},
  \bibinfo{person}{Laurent Charlin}, {and} \bibinfo{person}{David~M. Blei}.}
  \bibinfo{year}{2020}\natexlab{}.
\newblock \showarticletitle{Causal Inference for Recommender Systems}. In
  \bibinfo{booktitle}{\emph{RecSys}}. \bibinfo{publisher}{{ACM}},
  \bibinfo{pages}{426--431}.
\newblock


\bibitem[Wei et~al\mbox{.}(2021)]%
        {WeiFCWYH21}
\bibfield{author}{\bibinfo{person}{Tianxin Wei}, \bibinfo{person}{Fuli Feng},
  \bibinfo{person}{Jiawei Chen}, \bibinfo{person}{Ziwei Wu},
  \bibinfo{person}{Jinfeng Yi}, {and} \bibinfo{person}{Xiangnan He}.}
  \bibinfo{year}{2021}\natexlab{}.
\newblock \showarticletitle{Model-agnostic counterfactual reasoning for
  eliminating popularity bias in recommender system}. In
  \bibinfo{booktitle}{\emph{KDD}}. \bibinfo{pages}{1791--1800}.
\newblock


\bibitem[Zhang et~al\mbox{.}(2022)]%
        {bc_loss}
\bibfield{author}{\bibinfo{person}{An Zhang}, \bibinfo{person}{Wenchang Ma},
  \bibinfo{person}{Xiang Wang}, {and} \bibinfo{person}{Tat seng Chua}.}
  \bibinfo{year}{2022}\natexlab{}.
\newblock \showarticletitle{Incorporating Bias-aware Margins into Contrastive
  Loss for Collaborative Filtering}. In \bibinfo{booktitle}{\emph{{NeurIPS}}}.
\newblock


\bibitem[Zhang et~al\mbox{.}(2021)]%
        {ZhangFHWSLZ21}
\bibfield{author}{\bibinfo{person}{Yang Zhang}, \bibinfo{person}{Fuli Feng},
  \bibinfo{person}{Xiangnan He}, \bibinfo{person}{Tianxin Wei},
  \bibinfo{person}{Chonggang Song}, \bibinfo{person}{Guohui Ling}, {and}
  \bibinfo{person}{Yongdong Zhang}.} \bibinfo{year}{2021}\natexlab{}.
\newblock \showarticletitle{Causal Intervention for Leveraging Popularity Bias
  in Recommendation}.
\newblock \bibinfo{journal}{\emph{SIGIR}} (\bibinfo{year}{2021}),
  \bibinfo{pages}{11--20}.
\newblock


\bibitem[Zheng et~al\mbox{.}(2021)]%
        {ZhengGLHLJ21}
\bibfield{author}{\bibinfo{person}{Yu Zheng}, \bibinfo{person}{Chen Gao},
  \bibinfo{person}{Xiang Li}, \bibinfo{person}{Xiangnan He},
  \bibinfo{person}{Yong Li}, {and} \bibinfo{person}{Depeng Jin}.}
  \bibinfo{year}{2021}\natexlab{}.
\newblock \showarticletitle{Disentangling User Interest and Conformity for
  Recommendation with Causal Embedding}. In \bibinfo{booktitle}{\emph{{WWW}}}.
  \bibinfo{publisher}{{ACM} / {IW3C2}}, \bibinfo{pages}{2980--2991}.
\newblock


\bibitem[Zhu et~al\mbox{.}(2020)]%
        {ZhuHZC20}
\bibfield{author}{\bibinfo{person}{Ziwei Zhu}, \bibinfo{person}{Yun He},
  \bibinfo{person}{Yin Zhang}, {and} \bibinfo{person}{James Caverlee}.}
  \bibinfo{year}{2020}\natexlab{}.
\newblock \showarticletitle{Unbiased Implicit Recommendation and Propensity
  Estimation via Combinational Joint Learning}. In
  \bibinfo{booktitle}{\emph{RecSys}}. \bibinfo{pages}{551–556}.
\newblock


\end{thebibliography}

% \input{sec-appendix}
% \appendix

\end{document}